\documentclass[a4paper]{jpconf}
\pdfoutput=1 
\usepackage{graphicx}
\usepackage{booktabs} 
\usepackage{bbold}
\begin{document}
\title{Phenomenology of dark matter-nucleon effective interactions}

\author{Riccardo Catena}

\address{Chalmers University of Technology, Department of Physics, SE-412 96 G\"oteborg, Sweden}

\ead{catena@chalmers.se}

\begin{abstract}
I compare the non-relativistic effective theory of one-body dark matter-nucleon interactions to current dark matter direct detection experiments and neutrino telescope observations, presenting exclusion limits on the coupling constants of the theory.~In the analysis of direct detection experiments, I focus on the interference of different dark matter-nucleon interaction operators and on predictions observable at directional detectors.~Interpreting neutrino telescope observations, I use new nuclear response functions recently derived through nuclear structure calculations and show that hydrogen is not the most important element in the exclusion limit calculation for the majority of the spin-dependent dark matter-nucleon interaction operators.
\end{abstract}

\section{Introduction}
The solar system motion in the galactic rest frame induces a flux of Milky Way dark matter particles across the Earth and the Sun.~If dark matter interacts with nucleons, this flux is in principle detectable at direct detection experiments, at directional detectors, and, indirectly, at neutrino telescopes~\cite{Catena:2013pka}.  

Direct detection experiments search for nuclear recoil events induced by the scattering of Milky Way dark matter particles in low-background detectors.~The expected differential rate of recoil events per unit detector mass is 
\begin{equation}
\frac{{\rm d}\mathcal{R}}{{\rm d}E_{R}} = 
\sum_{T} \xi_T \frac{\rho_{\chi}}{m_\chi m_T}  
 \int_{u > v_{\rm min}(E_R)} \,  u\,f({\bf u} + {\bf v}_e(t))\, \frac{{\rm d}\sigma_{T}\left(u^2,E_R\right)}{{\rm d}E_R} \, d^3{\bf u} \,,
\label{rate_theory}
\end{equation}
where ${\rm d}\sigma_{T}(u^2,E_R)/{\rm d}E_R$ is the differential cross-section for dark matter scattering from nuclei of mass $m_T$ and mass fraction $\xi_T$,~$m_\chi$ is the dark matter particle mass, $\rho_\chi$ is the local dark matter density and ${\bf u}$ is the dark matter particle velocity in the detector rest frame.~The minimum velocity required to transfer a momentum $q=\sqrt{2 m_T E_R}$ in the scattering is given by $v_{\rm min}(E_R)=\sqrt{2 m_T E_R}/2\mu_T$, where $\mu_T$ is the dark matter-nucleus reduced mass.~The vector ${\bf v}_e(t)$ is the time-dependent Earth velocity in the galactic rest frame.~Here we consider a Maxwell-Boltzmann distribution $f({\bf u} + {\bf v}_e(t))\propto \exp(-|{\bf u}+{\bf v}_{e}(t)|^2/v_0^2)$ truncated at the local escape velocity $v_{\rm esc}=554$ km~s$^{-1}$, and $v_{0}=220$ km~s$^{-1}$~\cite{Catena:2009mf,Bozorgnia:2013pua,Catena:2011kv}.

Directional detectors are in a research and development stage.~They are designed to be sensitive to the momentum vector of nuclei recoiling against Milky Way dark matter particles.~The expected double differential rate of recoil events per unit detector mass is
\begin{eqnarray}
\frac{{\rm d}^2\mathcal{R}}{{\rm d}E_{R}\,{\rm d}\Omega} = \sum_{T} \frac{\xi_T}{(2\pi)} \frac{\rho_{\chi}}{m_\chi m_T}  
 \int \delta({\bf u}\cdot {\bf w}-v_{\rm min})\, u^2\,f({\bf u} + {\bf v_e}(t)) \, \frac{{\rm d}\sigma_{T}\left(u^2,E_R\right)}{{\rm d}E_R} \, {\rm d}^3{\bf u} \,,
\label{rate_theory2}
\end{eqnarray}
where ${\bf w}$ is the nuclear recoil direction.~Here we assume azimuthal symmetry around the direction of ${\bf v_e}(t)$, i.e. ${\rm d}\Omega=2\pi {\rm d}\hspace{-0.5mm}\cos\theta$.

Finally, neutrino telescopes search for energetic neutrinos produced by dark matter annihilation in the Sun.~Their flux depends on the rate of scattering from an initial dark matter particle velocity $w$ to a velocity less than the escape velocity $v(R)$ at a distance $R$ from the Sun's centre~\cite{Gould:1987ir}:
\begin{equation}
\Omega_{v}^{-}(w)= \sum_T n_T w\,\Theta\left( \frac{\mu_T}{\mu^2_{+,T}} - \frac{u^2}{w^2} \right)\int_{E u^2/w^2}^{E \mu_T/\mu_{+,T}^2} {\rm d}E_R\,\frac{{\rm d}\sigma_{T}\left(w^2,E_R\right)}{{\rm d}E_R}\,,
\label{eq:omega}
\end{equation}
where $E=m_\chi w^2/2$, is the initial dark matter particle kinetic energy, $n_T(R)$ is the density of the $T$ isotope in the Sun, and $u$ is the velocity of the dark matter particle at $R\rightarrow \infty$.~The dimensionless parameters $\mu_T$ and $\mu_{\pm,T}$  are defined as follows: $\mu_T\equiv m_\chi/m_T$ and $\mu_{\pm,T}\equiv (\mu_T\pm 1)/2$.~In the equations above, the differential cross-section for dark matter-nucleus scattering is commonly assumed to be
\begin{equation}
\frac{{\rm d}\sigma_{\rm T}(u^2,E_R)}{{\rm d}E_R} = \frac{2m_T}{(2J+1)u^2} \sum_{\tau,\tau'} \Bigg[ c_1^{\tau} c_1^{\tau'} W_{\rm SI}^{\tau\tau'}(E_R) + \frac{j_\chi(j_\chi+1)}{12} c_4^\tau c_4^{\tau'} W_{\rm SD}^{\tau\tau'}(E_R) \Bigg] \,, 
\label{eq:sigma_standard}
\end{equation}
where $W_{\rm SI}^{\tau\tau'}(E_R)=W_M^{\tau\tau'}(E_R)$ and $W_{\rm SD}^{\tau\tau'}(E_R)=W_{\Sigma'}^{\tau\tau'}(E_R)+W_{\Sigma''}^{\tau\tau'}(E_R)$ are nuclear response functions introduced below in Eq.~(\ref{eq:sigma}), $j_\chi$ and $J$ are the dark matter particle and nucleus spin, respectively, and $c_1^\tau$ and $c_4^\tau$, $\tau=0,1$, are coupling constants.~In the limit $E_R\rightarrow 0$, Eq.~(\ref{eq:sigma_standard}) can be written as ${\rm d}\sigma_{\rm T}/{\rm d}E_R=\left(\sigma_{SI}+\sigma_{SD}\right)/E_{\rm max}$, where $\sigma_{SI}$ and $\sigma_{SD}$ are the familiar spin-independent and spin-dependent cross-sections, respectively, and $E_{\rm max}=2 \mu_T^2 u^2/m_T$ is the maximum recoil energy.~Although a well motivated first approximation, Eq.~(\ref{eq:sigma_standard}) relies on a simplified picture of the actual complexity of dark matter-nucleon interactions.~The experimental efforts planned for the next years motivate the exploration of more advanced strategies. 

\begin{table}[t]
    \centering
    \begin{tabular}{ll}
    \toprule
        $\hat{\mathcal{O}}_1 = \mathbb{1}_{\chi N}$ & $\hat{\mathcal{O}}_9 = i{\hat{\bf{S}}}_\chi\cdot\left(\hat{{\bf{S}}}_N\times\frac{{\hat{\bf{q}}}}{m_N}\right)$  \\
        $\hat{\mathcal{O}}_3 = i\hat{{\bf{S}}}_N\cdot\left(\frac{{\hat{\bf{q}}}}{m_N}\times{\hat{\bf{v}}}^{\perp}\right)$ \hspace{2 cm} &   $\hat{\mathcal{O}}_{10} = i\hat{{\bf{S}}}_N\cdot\frac{{\hat{\bf{q}}}}{m_N}$   \\
        $\hat{\mathcal{O}}_4 = \hat{{\bf{S}}}_{\chi}\cdot \hat{{\bf{S}}}_{N}$ &   $\hat{\mathcal{O}}_{11} = i{\hat{\bf{S}}}_\chi\cdot\frac{{\hat{\bf{q}}}}{m_N}$   \\                                                                             
        $\hat{\mathcal{O}}_5 = i{\hat{\bf{S}}}_\chi\cdot\left(\frac{{\hat{\bf{q}}}}{m_N}\times{\hat{\bf{v}}}^{\perp}\right)$ &  $\hat{\mathcal{O}}_{12} = \hat{{\bf{S}}}_{\chi}\cdot \left(\hat{{\bf{S}}}_{N} \times{\hat{\bf{v}}}^{\perp} \right)$ \\                                                                                                                 
        $\hat{\mathcal{O}}_6 = \left({\hat{\bf{S}}}_\chi\cdot\frac{{\hat{\bf{q}}}}{m_N}\right) \left(\hat{{\bf{S}}}_N\cdot\frac{\hat{{\bf{q}}}}{m_N}\right)$ &  $\hat{\mathcal{O}}_{13} =i \left(\hat{{\bf{S}}}_{\chi}\cdot {\hat{\bf{v}}}^{\perp}\right)\left(\hat{{\bf{S}}}_{N}\cdot \frac{{\hat{\bf{q}}}}{m_N}\right)$ \\   
        $\hat{\mathcal{O}}_7 = \hat{{\bf{S}}}_{N}\cdot {\hat{\bf{v}}}^{\perp}$ &  $\hat{\mathcal{O}}_{14} = i\left(\hat{{\bf{S}}}_{\chi}\cdot \frac{{\hat{\bf{q}}}}{m_N}\right)\left(\hat{{\bf{S}}}_{N}\cdot {\hat{\bf{v}}}^{\perp}\right)$  \\
        $\hat{\mathcal{O}}_8 = \hat{{\bf{S}}}_{\chi}\cdot {\hat{\bf{v}}}^{\perp}$  & $\hat{\mathcal{O}}_{15} = -\left(\hat{{\bf{S}}}_{\chi}\cdot \frac{{\hat{\bf{q}}}}{m_N}\right)\left[ \left(\hat{{\bf{S}}}_{N}\times {\hat{\bf{v}}}^{\perp} \right) \cdot \frac{{\hat{\bf{q}}}}{m_N}\right] $ \\                                                                               
    \bottomrule
    \end{tabular}
    \caption{Complete set of non-relativistic operators considered in this study.} 
    \label{tab:operators}
\end{table}

\section{Effective theory of dark matter-nucleon interactions}
In this section I introduce the effective theory of dark matter-nucleon interactions~\cite{Fan:2010gt,Fitzpatrick:2012ix,Fitzpatrick:2012ib,Anand:2013yka}.~The focus is on showing how Eq.~(\ref{eq:sigma_standard}) generalizes in this context.~The effective theory of dark matter-nucleon interactions is based on the assumption: $|\mathbf{q}|/M_m\ll 1$, where $M_m$ is the mass scale of the exchanged mediators.~The theory is then constructed in terms of fundamental symmetries and basic operators.~The fundamental symmetry underlying the theory is Galilean invariance, which characterizes the non-relativistic nature of the interaction.~Simple kinematical arguments allow to identify five basic Galilean invariant operators, namely: $\mathbb{1}_{\chi N}$, ${\hat{\bf{q}}}$, $\hat{\bf{v}}^{\perp}$, ${\hat{\bf{S}}}_N$, and ${\hat{\bf{S}}}_\chi$, where $\mathbb{1}_{\chi N}$ is the identity, ${\hat{\bf{q}}}$ is the momentum transfer operator, $\hat{\bf{v}}^{\perp}$ is the transverse relative velocity operator, and ${\hat{\bf{S}}}_N$ and ${\hat{\bf{S}}}_\chi$ are the nucleon and dark matter particle spin operators, respectively.~The most general Hamiltonian density for dark matter-nucleon interactions ${\bf\hat{\mathcal{H}}}({\bf{r}})$ is then constructed as an expansion in powers of $\hat{\mathbf{q}}/M_m$ from the basic operators.~Truncating the expansion at second order in $\hat{\mathbf{q}}$, and assuming mediators of spin less or equal to 1, one finds that ${\bf\hat{\mathcal{H}}}({\bf{r}})$ is the linear combination of 14 Galilean invariant operators:
\begin{equation}
{\bf\hat{\mathcal{H}}}({\bf{r}}) = \sum_{\tau=0,1} \sum_{k} c_k^{\tau} \,\hat{\mathcal{O}}_{k}({\bf{r}}) \, t^{\tau} \,.
\label{eq:Hc0c1}
\end{equation}
The 14 operators $\hat{\mathcal{O}}_{k}$ in Eq.~(\ref{eq:Hc0c1}) are listed in Tab.~\ref{tab:operators}.~$\hat{\mathcal{O}}_1$ and $\hat{\mathcal{O}}_4$ correspond to the familiar spin-independent and spin-dependent interactions, respectively.~Isoscalar and isovecotor coupling constants in~(\ref{eq:Hc0c1}), $c_k^{0}$ and $c_k^{1}$ respectively, are related to the coupling constants for protons ($c^{p}_k$) and neutrons ($c^{n}_k$) as follows: $c^{p}_k=(c^{0}_k+c^{1}_k)/2$, and $c^{n}_k=(c^{0}_k-c^{1}_k)/2$.~These constants have dimension mass to the power $-2$.~The differential cross-section for dark matter-nucleus scattering follows from Eq.~(\ref{eq:Hc0c1}): 
\begin{eqnarray}
\frac{{\rm d} \sigma_{T}(u^2,E_R)}{{\rm d} E_R} = \frac{2m_{T}}{u^2} \frac{1}{2J+1} \sum_{\tau,\tau'} &\Bigg[& \sum_{k=M,\Sigma',\Sigma''} R^{\tau\tau'}_k\left(u_T^{\perp 2}, {q^2 \over m_N^2} \right) W_k^{\tau\tau'}(E_R) \nonumber\\
&+&{q^{2} \over m_N^2} \sum_{k=\Phi'', \Phi'' M, \tilde{\Phi}', \Delta, \Delta \Sigma'} R^{\tau\tau'}_k\left(v_T^{\perp 2}, {q^2 \over m_N^2}\right) W_k^{\tau\tau'}(E_R) \Bigg] \nonumber\\ \,,
\label{eq:sigma}
\end{eqnarray}
where $q^2=2 m_T E_R$, $v_T^{\perp 2}=u^2-q^2/{(4 \mu_T^2})$ and $m_N$ is the nucleon mass.~The dark matter response functions $R^{\tau\tau'}_k$ are known analytically and depend on the momentum transfer, the dark matter-nucleus relative velocity $u$ and the coupling constants $c_k^\tau$.~The isotope dependent nuclear response functions $W_k^{\tau\tau'}$ must be computed numerically through nuclear-structure calculations.~For Xe, Ge, Na, I and F, I use the response functions found in~\cite{Anand:2013yka}.~For the 16 most abundant elements in the Sun, I adopt the functions $W_k^{\tau\tau'}$ published in~\cite{Catena:2015uha}.~From Eqs.~(\ref{eq:sigma_standard}), (\ref{eq:sigma}) and Tab.~\ref{tab:operators}, the contrast between the standard paradigm based on the familiar spin-independent and spin-dependent interactions and the actual complexity of dark matter-nucleon interactions is manifest. 

\begin{figure}[t]
\begin{center}
\begin{minipage}[t]{0.32\linewidth}
\vspace{-3.82 cm}
\centering
\includegraphics[width=0.89\textwidth]{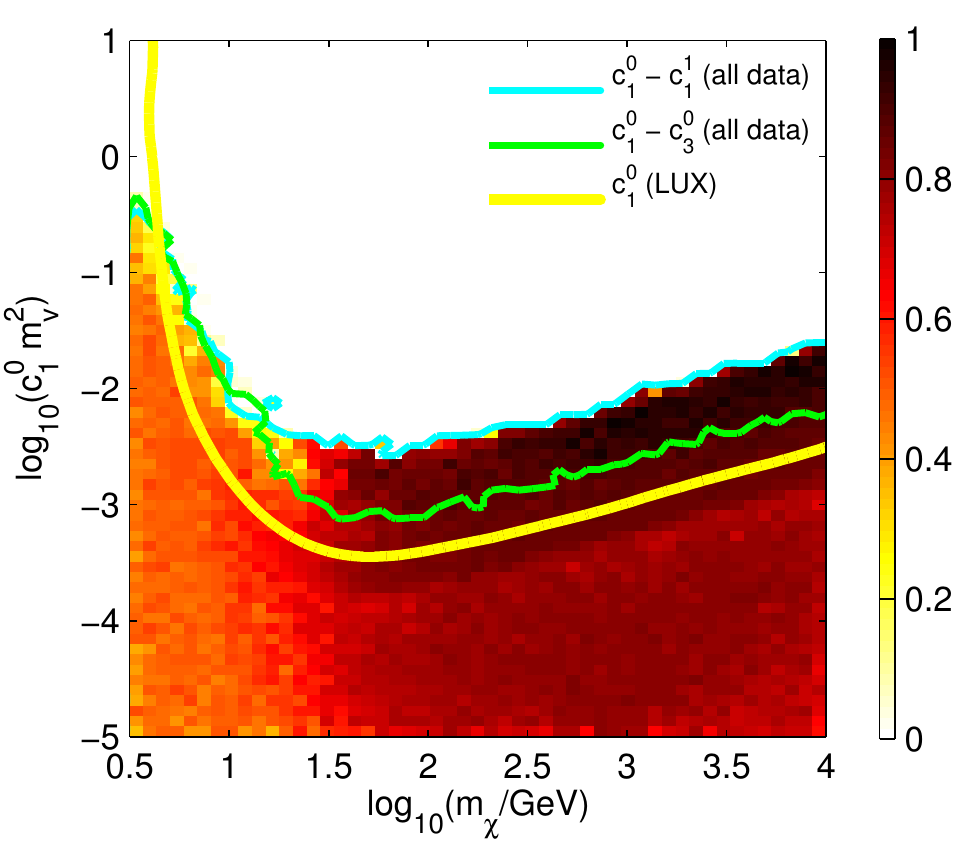}
\end{minipage}
\begin{minipage}[t]{0.32\linewidth}
\centering
\includegraphics[width=\textwidth]{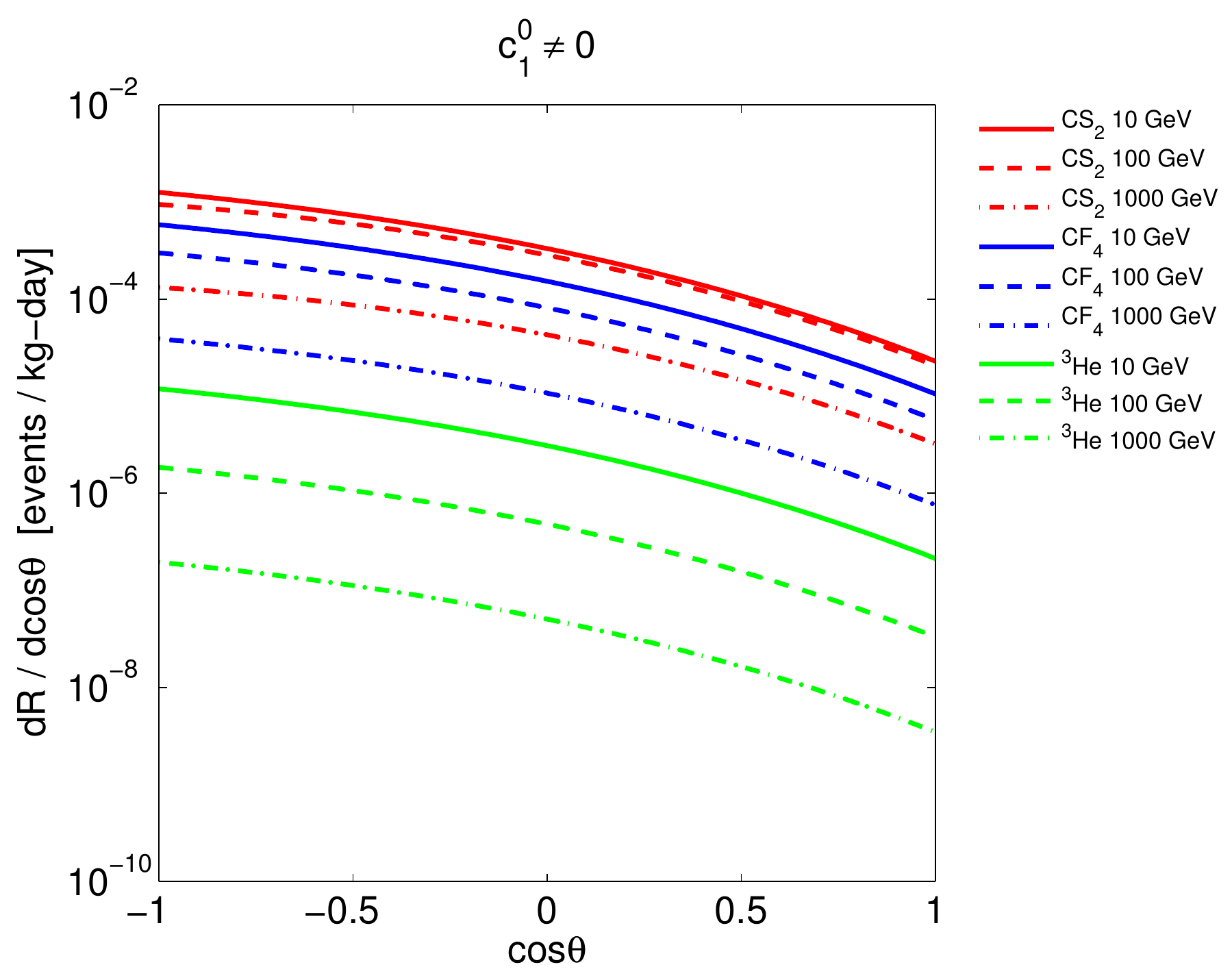}
\end{minipage}
\begin{minipage}[t]{0.32\linewidth}
\centering
\includegraphics[width=\textwidth]{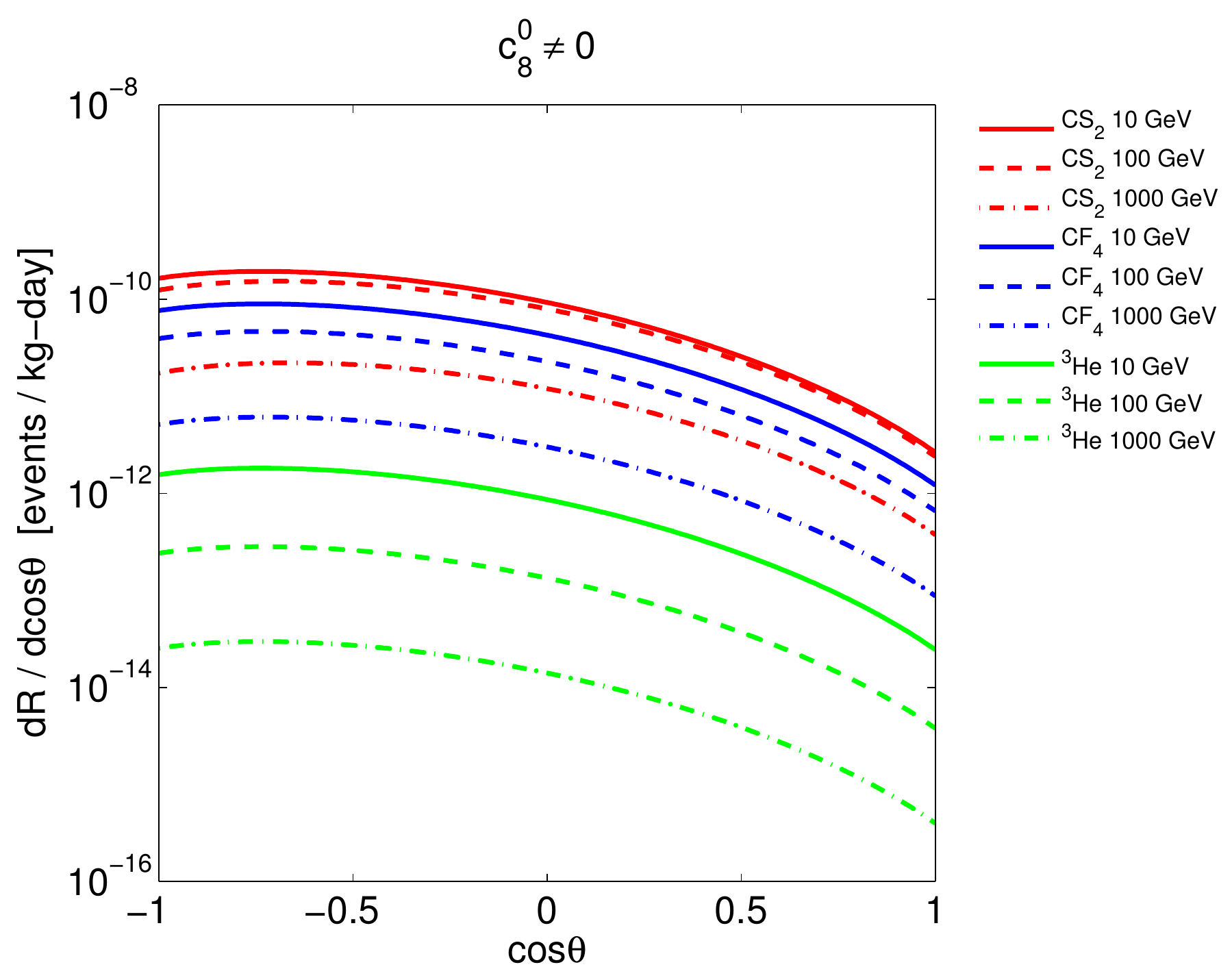}
\end{minipage}
\label{fig:dd-ddd}
\caption{{\it Left panel}.~Exclusion limits on $c_1^0$.~The yellow line considers LUX only and neglects interference effects.~Green and cyan lines consider different experiments simultaneously and include $c_1^0-c_3^0$ and $c_1^0-c_1^1$ interference effects, respectively.~{\it Central panel.}~Directional differential rate as a function of $\cos\theta$ for $\hat{\mathcal{O}}_1$.~{\it Right panel.}~Same as for the central panel, but now for $\hat{\mathcal{O}}_8$.}
\end{center}
\end{figure}

\section{Phenomenology}
Having introduced the effective theory of dark matter-nucleon interactions, I now compare its predictions to current observations.

\subsection{Direct detection experiments}
Comparing the effective theory of dark matter-nucleon interactions to current direct detection data, I have identified two main results.~Firstly, current data can place interesting constraints on the coupling constants of interaction operators commonly neglected.~For instance, limits on the strength of the interactions $\hat{\mathcal{O}}_8$ and $\hat{\mathcal{O}}_{11}$ are comparable with those found for the familiar spin-dependent operator $\hat{\mathcal{O}}_4$~\cite{Catena:2014uqa,Catena:2015uua}.~Secondly, operator interference plays an important role in the calculation of exclusion limits on the coupling constants $c_k^\tau$.~Specifically, I found that destructive interference effects can weaken standard direct detection exclusion limits by up to one order of magnitude in the coupling constants~\cite{Catena:2015uua}.~This conclusion is illustrated in the left panel of Fig.~1.

\subsection{Directional detectors}
Integrating Eq.~(\ref{rate_theory2}) over all recoil energies, one obtains the differential rate of recoil events per unit detector mass ${\rm d}\mathcal{R}/{\rm d}\cos\theta$.~The central and right panels in Fig.~1 show ${\rm d}\mathcal{R}/{\rm d}\cos\theta$ as a function of $\cos\theta$ for the operators $\hat{\mathcal{O}}_1$ and $\hat{\mathcal{O}}_8$, respectively.~For $\hat{\mathcal{O}}_1$ the majority of the recoil events is expected in a direction opposite to the Earth's motion, since the maximum of ${\rm d}\mathcal{R}/{\rm d}\cos\theta$ is at $\cos\theta=-1$.~In contrast, for velocity dependent operators like $\hat{\mathcal{O}}_8$, ${\rm d}\mathcal{R}/{\rm d}\cos\theta$ is maximum at $\cos\theta>-1$ and the majority of the recoil events is expected in a ring around $-{\bf v}_e$.~Such interaction dependent ring-like structures in the sphere of recoil directions can be used to discriminate among different dark matter-nucleon interactions~\cite{Catena:2015vpa,Kavanagh:2015jma}.

\subsection{Neutrino telescopes}
Below I highlight three key results found studying the phenomenology of dark matter-nucleon interactions at neutrino telescopes~\cite{Catena:2015uha,Catena:2015iea}:
\begin{enumerate}
\item The operator $\hat{\mathcal{O}}_{11}$ induces a rate of dark matter capture by the Sun which for $m_\chi>30$~GeV is larger than the rate associated with $\hat{\mathcal{O}}_{4}$.~In fact, inspection of Eq.~(\ref{eq:sigma}) shows that $\hat{\mathcal{O}}_{11}$ generates the same nuclear response function of $\hat{\mathcal{O}}_{1}$, which is generically larger than the one associated with the spin-dependent operator $\hat{\mathcal{O}}_{4}$ (see left panel in Fig.~2).
\item For velocity dependent operators like $\hat{\mathcal{O}}_7$, neutrino telescopes are superior to current direct detection experiments, since dark matter particles move faster in the Sun than on Earth, as they gain speed crossing the Sun's gravitational potential (see central panel in Fig.~2).
\item Hydrogen is not the most important element in the exclusion limit calculation for the majority of the spin-dependent operators in Tab.~\ref{tab:operators}.~Although less abundant, heavier elements produce a broader range of recoil energies than Hydrogen, and hence a larger capture rate for differential cross-sections going to zero in the limit $E_R\rightarrow 0$ (see right panel in Fig.~2, and Eq.~(\ref{eq:omega})).
\end{enumerate}

\begin{figure}[t]
\begin{center}
\begin{minipage}[t]{0.32\linewidth}
\centering
\includegraphics[width=\textwidth]{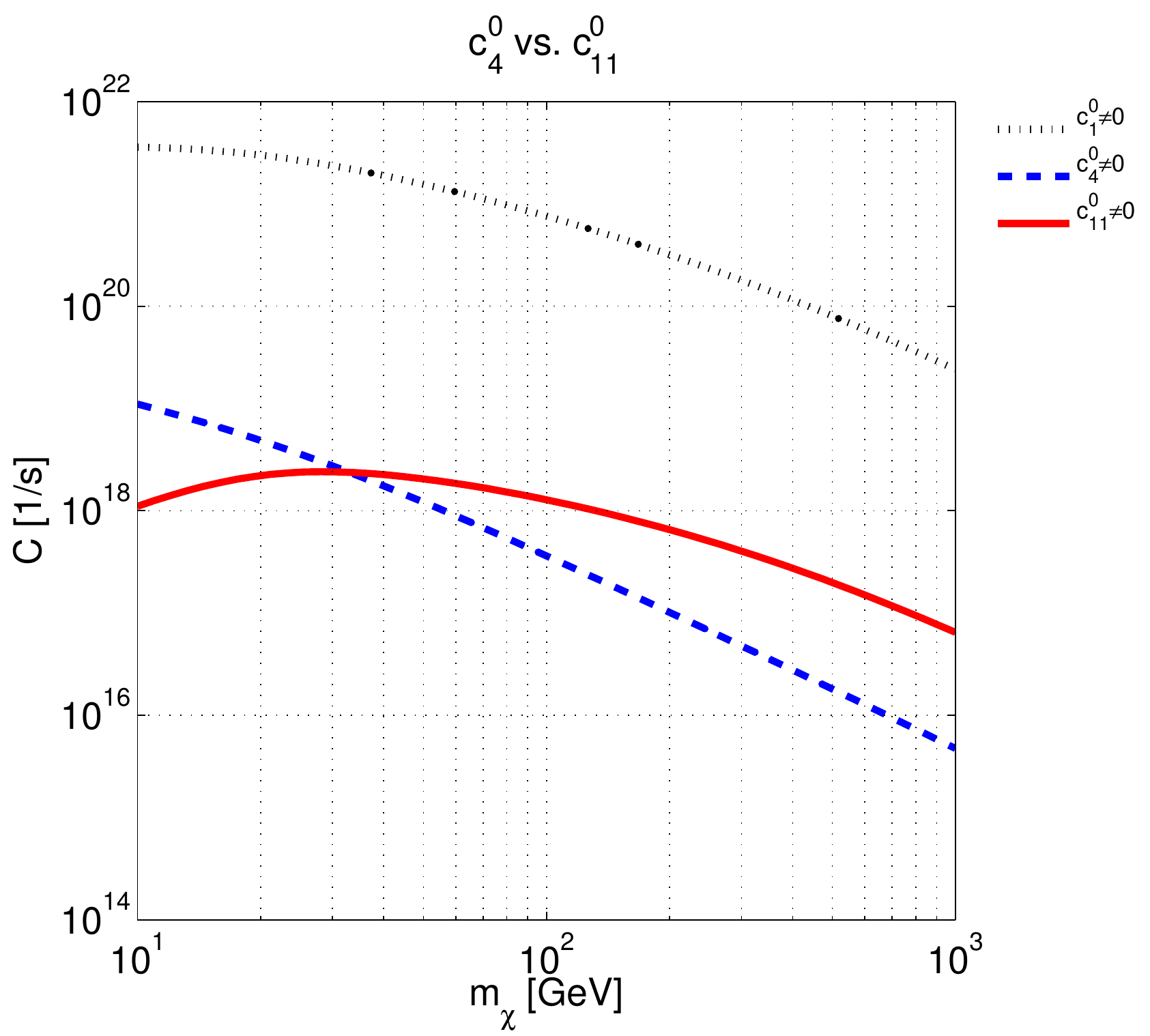}
\end{minipage}
\begin{minipage}[t]{0.32\linewidth}
\centering
\includegraphics[width=0.871\textwidth]{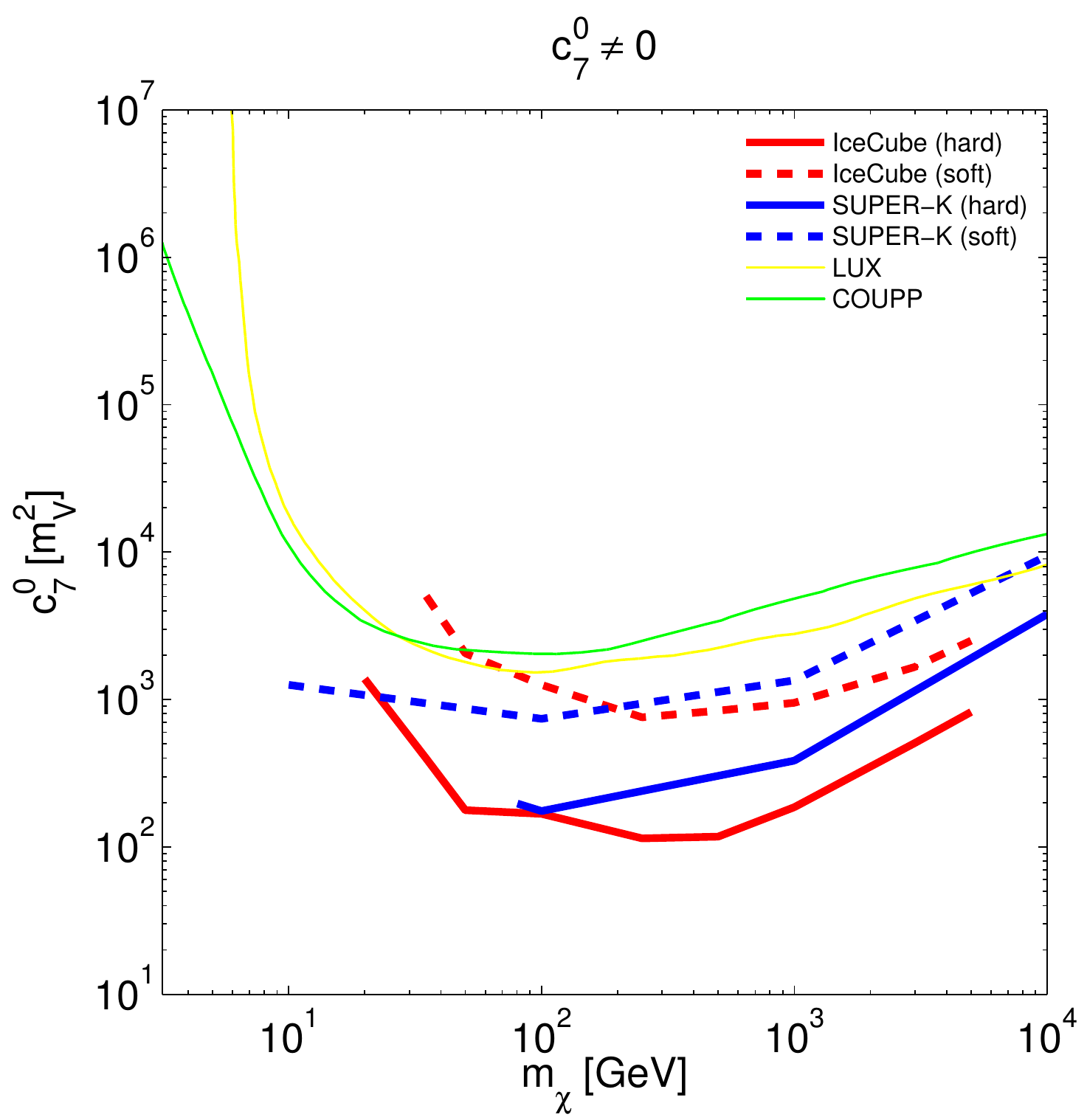}
\end{minipage}
\begin{minipage}[t]{0.32\linewidth}
\centering
\includegraphics[width=\textwidth]{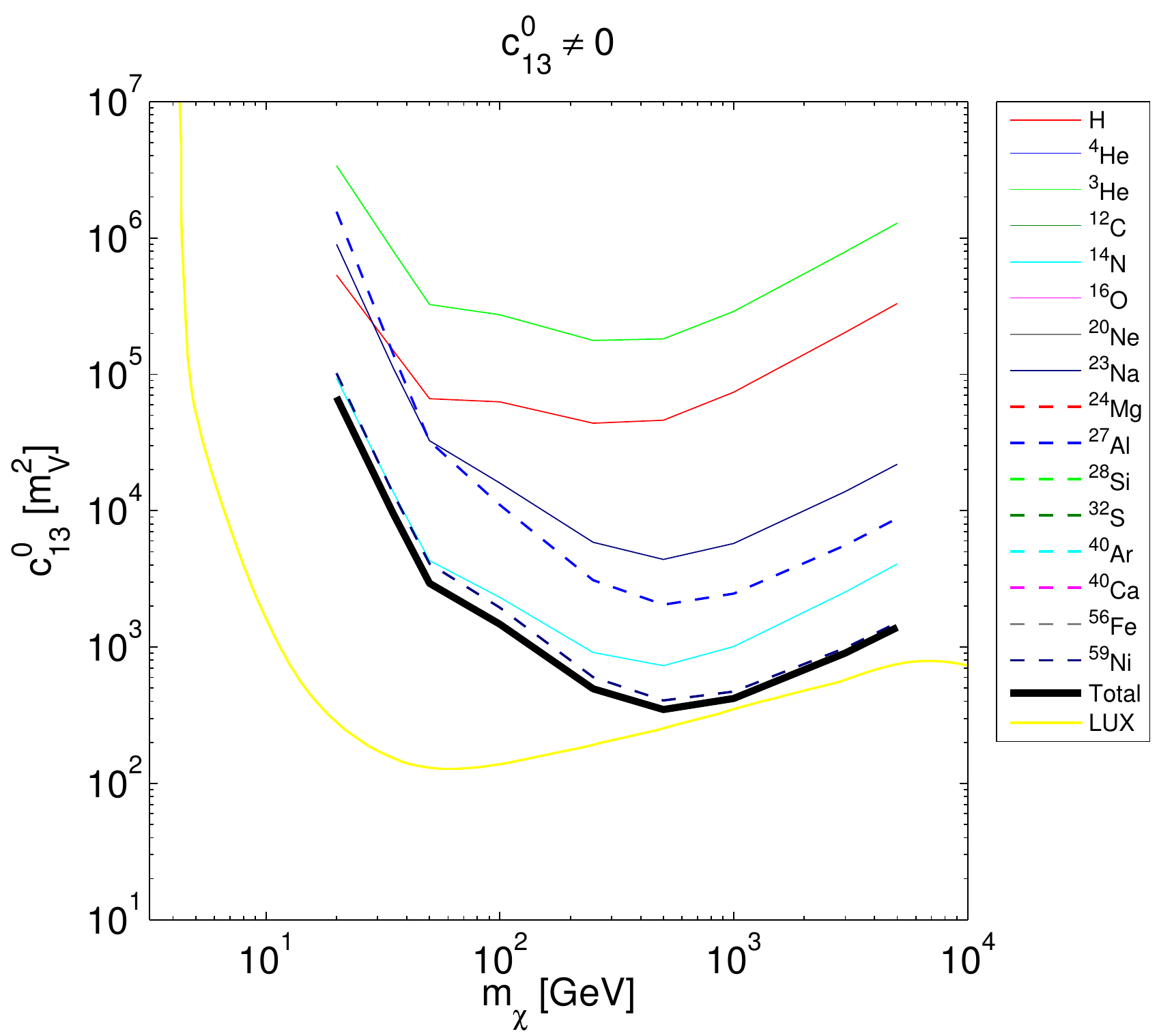}
\end{minipage}
\label{fig:nt}
\caption{{\it Left Panel.}~Capture rate as a function of $m_\chi$ for the operators $\hat{\mathcal{O}}_1$ (dotted line), $\hat{\mathcal{O}}_4$ (dashed line) and $\hat{\mathcal{O}}_{11}$ (solid line).~{\it Central panel.}~Exclusion limits on $c_7^0$ from the experiments in the legend.~{\it Right panel.}~Exclusion limits on $c_{13}^0$ from IceCube (hard).~In the capture by the Sun, dark matter is assumed to scatter on single elements, as shown in the legend.}
\end{center}
\end{figure}

\section{Conclusion}
The effective theory of dark matter-nucleon interactions is a promising framework for modeling the scattering of dark matter from nuclei.~From the theoretical perspective, its phenomenology is currently under investigation, e.g.~\cite{Fan:2010gt,Fitzpatrick:2012ix,Fitzpatrick:2012ib,Anand:2013yka,DelNobile:2013sia,Peter:2013aha,Hill:2013hoa,Catena:2014uqa,Catena:2014hla,Catena:2014epa,Gluscevic:2014vga,Panci:2014gga,Vietze:2014vsa,Barello:2014uda,Schneck:2015eqa,Catena:2015uua,Gluscevic:2015sqa,DelNobile:2015lxa,DelNobile:2015rmp} and \cite{Liang:2013dsa,Blumenthal:2014cwa,Vincent:2014jia,Catena:2015iea,Catena:2015vpa,Kavanagh:2015jma}.~Importantly, present observations can probe many of the dark matter-nucleon interactions which are currently neglected within the standard paradigm based on the familiar spin-independent and spin-dependent interactions. 

\section*{References}
\bibliography{ref}{}

\providecommand{\newblock}{}
\begin{thebibliography}{10}
\expandafter\ifx\csname url\endcsname\relax
  \def\url#1{{\tt #1}}\fi
\expandafter\ifx\csname urlprefix\endcsname\relax\def\urlprefix{URL }\fi
\providecommand{\eprint}[2][]{\url{#2}}

\bibitem{Catena:2013pka}
Catena R and Covi L 2014 {\em Eur.Phys.J.\/} {\bf C74} 2703 (\textit{Preprint}
  \eprint{1310.4776})

\bibitem{Catena:2009mf}
Catena R and Ullio P 2010 {\em JCAP\/} {\bf 1008} 004 (\textit{Preprint}
  \eprint{0907.0018})

\bibitem{Bozorgnia:2013pua}
Bozorgnia N, Catena R and Schwetz T 2013 {\em JCAP\/} {\bf 1312} 050
  (\textit{Preprint} \eprint{1310.0468})

\bibitem{Catena:2011kv}
Catena R and Ullio P 2012 {\em JCAP\/} {\bf 1205} 005 (\textit{Preprint}
  \eprint{1111.3556})

\bibitem{Gould:1987ir}
Gould A 1987 {\em Astrophys.J.\/} {\bf 321} 571

\bibitem{Fan:2010gt}
Fan J, Reece M and Wang L~T 2010 {\em JCAP\/} {\bf 1011} 042 (\textit{Preprint}
  \eprint{1008.1591})

\bibitem{Fitzpatrick:2012ix}
Fitzpatrick A~L, Haxton W, Katz E, Lubbers N and Xu Y 2013 {\em JCAP\/} {\bf
  1302} 004 (\textit{Preprint} \eprint{1203.3542})

\bibitem{Fitzpatrick:2012ib}
Fitzpatrick A~L, Haxton W, Katz E, Lubbers N and Xu Y 2012  (\textit{Preprint}
  \eprint{1211.2818})

\bibitem{Anand:2013yka}
Anand N, Fitzpatrick A~L and Haxton W 2014 {\em Phys.Rev.\/} {\bf C89} 065501
  (\textit{Preprint} \eprint{1308.6288})

\bibitem{Catena:2015uha}
Catena R and Schwabe B 2015 {\em JCAP\/} {\bf 1504} 042 (\textit{Preprint}
  \eprint{1501.03729})

\bibitem{Catena:2014uqa}
Catena R and Gondolo P 2014 {\em JCAP\/} {\bf 1409} 045 (\textit{Preprint}
  \eprint{1405.2637})

\bibitem{Catena:2015uua}
Catena R and Gondolo P 2015 {\em JCAP\/} {\bf 1508} 022 (\textit{Preprint}
  \eprint{1504.06554})

\bibitem{Catena:2015vpa}
Catena R 2015 {\em JCAP\/} {\bf 1507} 026 (\textit{Preprint}
  \eprint{1505.06441})

\bibitem{Kavanagh:2015jma}
Kavanagh B~J 2015 {\em Phys. Rev.\/} {\bf D92} 023513 (\textit{Preprint}
  \eprint{1505.07406})

\bibitem{Catena:2015iea}
Catena R 2015 {\em JCAP\/} {\bf 1504} 052 (\textit{Preprint}
  \eprint{1503.04109})

\bibitem{DelNobile:2013sia}
Cirelli M, Del~Nobile E and Panci P 2013 {\em JCAP\/} {\bf 1310} 019
  (\textit{Preprint} \eprint{1307.5955})

\bibitem{Peter:2013aha}
Peter A~H, Gluscevic V, Green A~M, Kavanagh B~J and Lee S~K 2014 {\em Phys.Dark
  Univ.\/} {\bf 5-6} 45--74 (\textit{Preprint} \eprint{1310.7039})

\bibitem{Hill:2013hoa}
Hill R~J and Solon M~P 2014 {\em Phys.Rev.Lett.\/} {\bf 112} 211602
  (\textit{Preprint} \eprint{1309.4092})

\bibitem{Catena:2014hla}
Catena R 2014 {\em JCAP\/} {\bf 1409} 049 (\textit{Preprint}
  \eprint{1407.0127})

\bibitem{Catena:2014epa}
Catena R 2014 {\em JCAP\/} {\bf 1407} 055 (\textit{Preprint}
  \eprint{1406.0524})

\bibitem{Gluscevic:2014vga}
Gluscevic V and Peter A~H~G 2014 {\em JCAP\/} {\bf 1409} 040 (\textit{Preprint}
  \eprint{1406.7008})

\bibitem{Panci:2014gga}
Panci P 2014 {\em Adv.High Energy Phys.\/} {\bf 2014} 681312 (\textit{Preprint}
  \eprint{1402.1507})

\bibitem{Vietze:2014vsa}
Vietze L, Klos P, Menéndez J, Haxton W and Schwenk A 2014  (\textit{Preprint}
  \eprint{1412.6091})

\bibitem{Barello:2014uda}
Barello G, Chang S and Newby C~A 2014 {\em Phys.Rev.\/} {\bf D90} 094027
  (\textit{Preprint} \eprint{1409.0536})

\bibitem{Schneck:2015eqa}
Schneck K, Cabrera B, Cerdeno D, Mandic V, Rogers H {\em et~al.\/} 2015
  (\textit{Preprint} \eprint{1503.03379})

\bibitem{Gluscevic:2015sqa}
Gluscevic V, Gresham M~I, McDermott S~D, Peter A~H~G and Zurek K~M 2015
  (\textit{Preprint} \eprint{1506.04454})

\bibitem{DelNobile:2015lxa}
Del~Nobile E, Gelmini G~B, Georgescu A and Huh J~H 2015 {\em JCAP\/} {\bf 1508}
  046 (\textit{Preprint} \eprint{1502.07682})

\bibitem{DelNobile:2015rmp}
Del~Nobile E, Gelmini G~B and Witte S~J 2015  (\textit{Preprint}
  \eprint{1512.03961})

\bibitem{Liang:2013dsa}
Liang Z~L and Wu Y~L 2014 {\em Phys.Rev.\/} {\bf D89} 013010 (\textit{Preprint}
  \eprint{1308.5897})

\bibitem{Blumenthal:2014cwa}
Blumenthal J, Gretskov P, Kr{\"a}mer M and Wiebusch C 2015 {\em Phys.Rev.\/}
  {\bf D91} 035002 (\textit{Preprint} \eprint{1411.5917})

\bibitem{Vincent:2014jia}
Vincent A~C, Scott P and Serenelli A 2015 {\em Phys.Rev.Lett.\/} {\bf 114}
  081302 (\textit{Preprint} \eprint{1411.6626})

\end{thebibliography}
\bibliographystyle{iopart-num}

\end{document}